\documentclass{article}
\usepackage{frascatiphys}
\begin{document}        
\title{STUDY OF THE ANGULAR RESOLUTION OF THE ARGO-YBJ DETECTOR}
\author{
Giuseppe Di Sciascio, Elvira Rossi on behalf of the ARGO-YBJ Collaboration    \\
{\em Dip. di Fisica Universit\`a di Napoli and INFN, sez. di Napoli, Italy} \\
}
\maketitle
\baselineskip=11.6pt
\begin{abstract}
The determination of the arrival direction of gamma-induced air showers in the ARGO-YBJ 
experiment has been investigated using different algorithms in the reconstruction procedure.
The calculation has been performed, as a function of pad multiplicity, for different primary 
energies. 
The performance of a conical correction to the shower front with a suitable fixed cone slope 
is discussed.
\end{abstract}
\baselineskip=14pt

\section{Introduction}
The ARGO-YBJ detector, currently under construction at the Yangbajing 
Laboratory (P.R. China, 4300 m a.s.l.), is a full coverage array of dimensions 
$\sim 74\times 78~m^2$ realized with a single layer of RPCs. The area 
surrounding this central detector ({\it carpet}), up to $\sim$ 100 $\times$ 110 $m^2$, 
is partially ($\sim 50 \%$) instrumented with RPCs.
The basic element is the logical {\it pad} ($56\times 62~cm^2$) which defines 
the time and space granularity of the detector. 

In a search for cosmic point-like sources the main problem is the background due to charged cosmic
rays, so a good angular resolution (measurement of the arrival direction) is necessary. 
In order to determine the primary direction, the shower front has to be reconstructed 
from measurement of the time and position of the fired detectors.
This calculation is usually performed for internal events only. In fact, the direction of
showers with the core outside the detector is in general badly reconstructed due to the cone-like
shape of the shower front and the unknown core position.

\begin{figure}[t]
\begin{minipage}[t]{.48\linewidth}
   \vspace{5.8cm}
\includegraphics{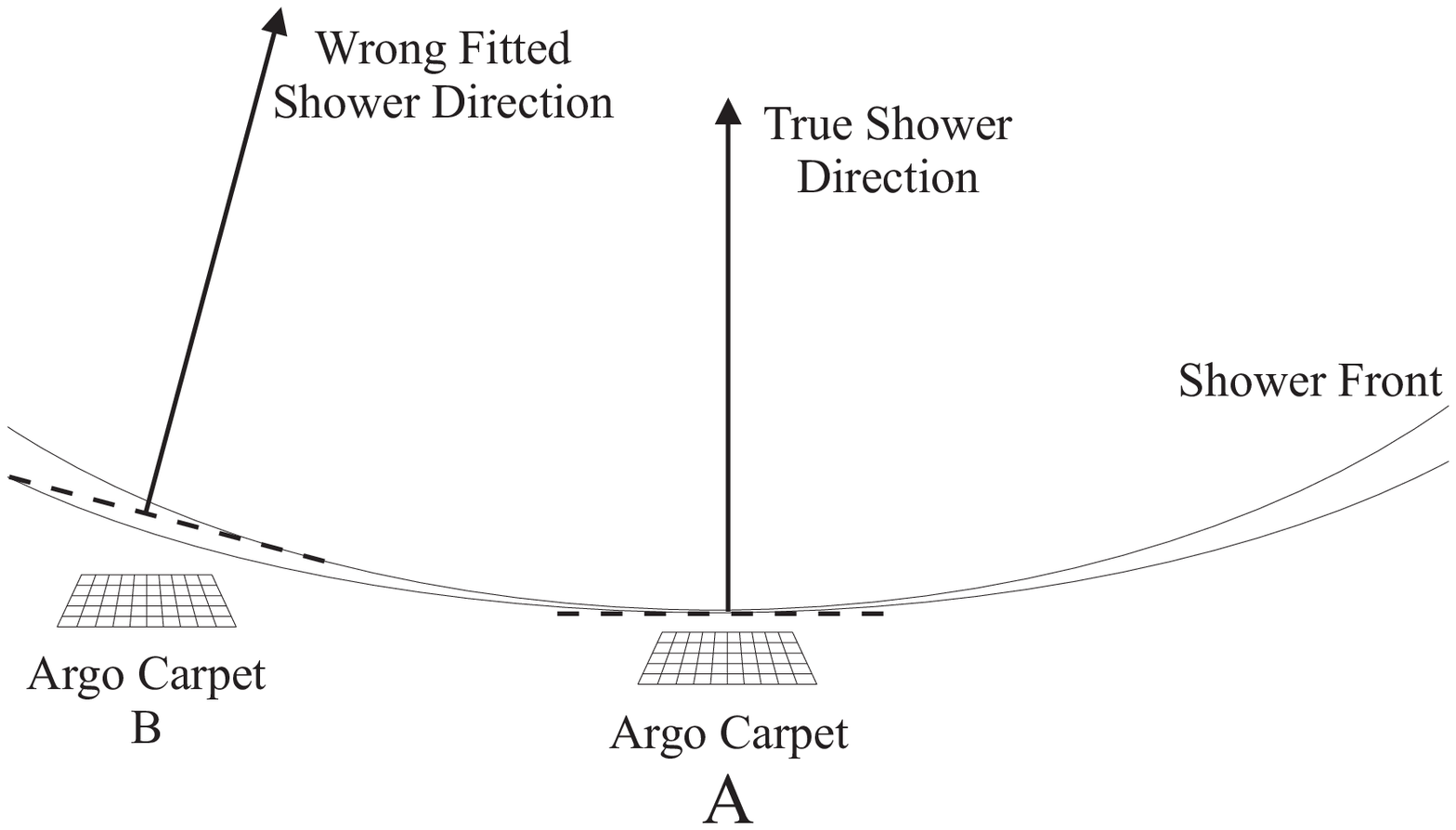}
 \caption{\it Core inside or outside the carpet can make the difference. The true shower
core hits under the central arrow.
    \label{misangle} }
 \end{minipage}\hfill     
\begin{minipage}[t]{.48\linewidth}
   \vspace{5.8cm}
    \includegraphics{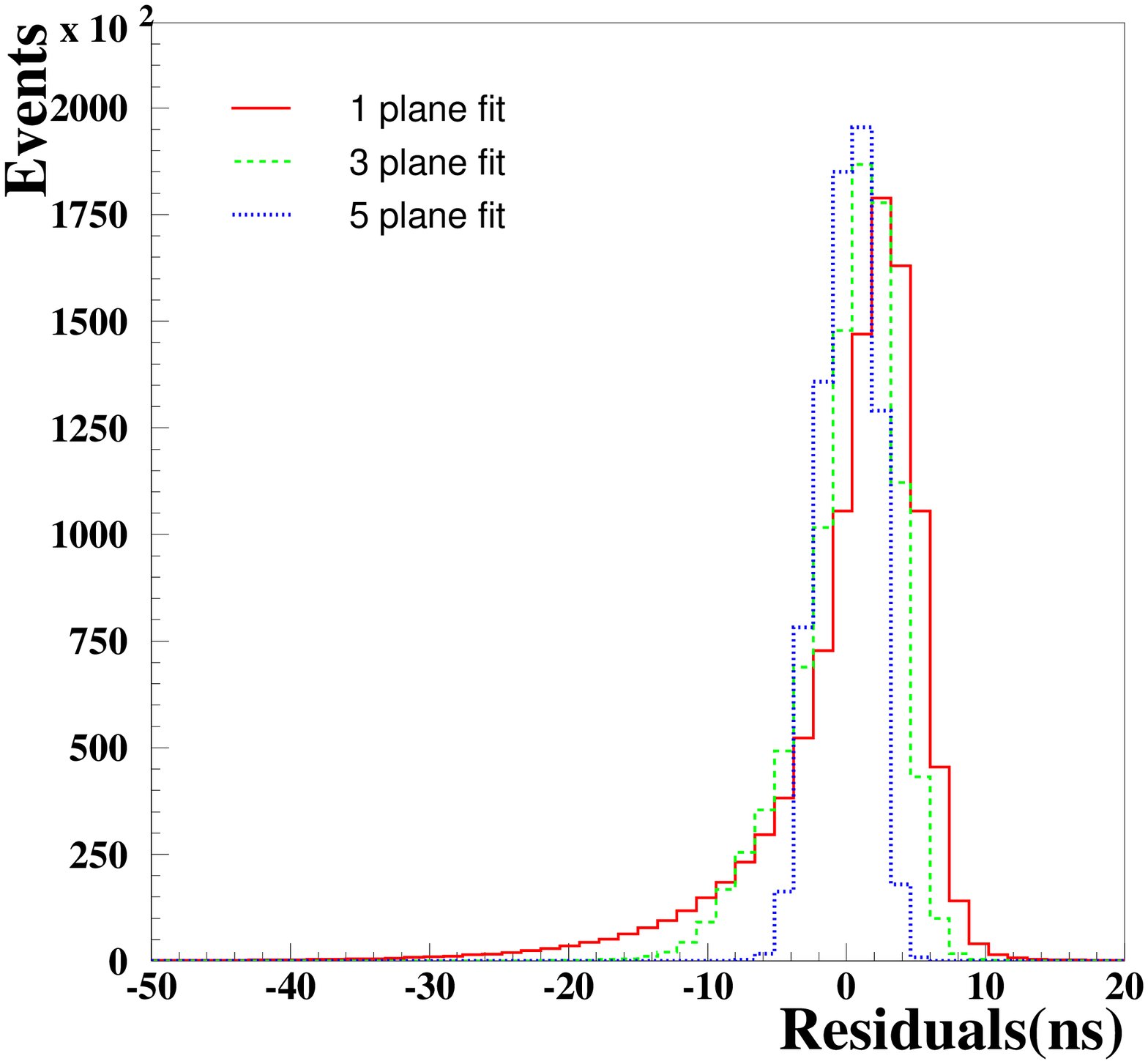}
    \caption{\it  Distributions of temporal residuals calculated for 1 TeV 
vertical $\gamma$-induced showers after a different number of plane fit iterations.
 \label{residuo1} }
 \end{minipage}\hfill        
\end{figure}

In Fig. \ref{misangle} is shown an exaggerated shower with two scenarios: case A where the
shower core strikes at the center of the ARGO-YBJ detector, and case B where the core hits outside 
the carpet. The primary energy in case B must be larger than that in case A to be able to trigger
the array with the same detector response (number of fired pads). In case A, the direction 
obtained from the plane fitting is close to the true shower direction. On the other hand, 
in case B,
the carpet sees a tilted portion of the same air shower, and in turn the fitted direction
acquires a pointing error and deviates from the true shower direction. 

In this paper we present different algorithms to reconstruct the shower direction. 
The calculations have been performed for different primary energies.
The effect of a lead converter sheet on the detector is also investigated.
The performance of a conical correction to the shower front with a suitable fixed cone slope 
is discussed.

\section{Reconstruction procedure of the shower direction}

The arrival time of the earliest particles provides the information concerning
the time profile of the shower front. From MC simulations (see Fig. 15 of ref.\cite{EPAS}) it
appears that the shower front assumes a parabolic shape more pronounced for the electron component
than for the photon component. Near the core charged particles form, at a first approximation, 
a plane disk. 

The thickness of an EAS is defined by the distribution of arrival times of the shower particles 
which are delayed relative to the prompt particles which make up the shower front. Therefore, 
the shower thickness is determined by the lower-energy particles in a shower. This means that the
shower thickness is the smallest near the core and increases with the increasing distance from the 
shower axis position\cite{EPAS}.
Hence, even in the case of vertical incidence the particles arrivals are not simultaneous.

In addition to the curvature, the combined effects of the finite size and time resolution 
of the detector make the shower apparently non-planar: since the shower plane has a 
thickness and the arrival time of the first particle is measured by the TDCs, 
the larger the number of particles, the earlier the measurement (on average).

The usual method for the reconstruction of the shower direction is a $\chi^2$ fit to the
recorded arrival times $t_i$ by minimization of 
\begin{equation}
 \chi^2 = \sum_i w ( f - t_i )^2 
\end{equation}
where the sum includes all pads with a time signal. Usually the function $f$ describes a plane,
a cone with a fixed cone slope, or a plane with curvature corrections depending on $r$ and
$N$. The weights $w$ are generally chosen to be an empirical function of the number $N$ of 
particles registered in a detector.

An improvement to this scheme can be achieved by excluding from the analysis the time values 
belonging to the non-gaussian tails of the arrival time distributions with
successive $\chi^2$ minimizations for each shower\cite{fititer}.

In this paper we will investigate the performance of the $\chi^2$ fit reconstructing the primary
direction of the showers sampled by the ARGO-YBJ carpet by means of the following iterative 
procedure:
\begin{itemize}
\item[{\bf a)}] {\bf Planar fit}\\
In the first step we recover the shower direction cosines $\{l_p,m_p\}$ by means of an 
unweighted planar fit performed minimizing the function 
\begin{equation}
\chi^2 (ns^2) = \frac{1}{c^2}\sum_i \{lx_i+my_i+nz_i+c(t_i-t_0)\}^2
\end{equation}
The sum includes all pads with a time signal $t_i$, $c$ is the velocity of light,
$(x_i,y_i,z_i)$ are the coordinates of the central position of the $i$-th pad. 
The parameters of the fit are the time offset $t_0$ and the direction 
cosines $l,m$. 
After each minimization the time signals which deviate more than $K\cdot \sigma$ from the 
fitting function are excluded from further analysis and the fit is iterated until all times 
do not verify this condition or the remaining hits number is $\leq 5$ (in this case the event is 
discarded). Here $\sigma$ is the standard deviation of the time distribution around
the fitted plane (i.e., $\sigma = \sqrt{\frac {\chi^2}{N-3}}$). 
In principle, the value of $K$ could depend on the features of the reconstructed showers 
as well as on the experimental condition (pad dimension, shaping of the signals, 
multihit capability, etc.).
\item[{\bf b)}] {\bf Conical correction}\\
With this first determination of the arrival direction  $\{l_p,m_p\}$ we calculate the conical 
correction
\begin{equation}
\delta t_i = \frac{\alpha}{c}\cdot R_i
\end{equation}
where
\begin{equation}
 R_i = \sqrt { (\Delta x_i^2 + \Delta y_i^2) - (\Delta x_i l_p + \Delta y_i m_p)^2 } 
\end{equation}
and $\Delta x_i = x_i - x_c$, $\Delta y_i = y_i - y_c$ are the pad distances from the shower core
position $\{x_c,y_c\}$.
Then we correct the experimental time values $t_i$
\begin{equation}
 t_i \to t'_i = t_i - \delta t_i(l_p,m_p) 
\end{equation}
In this approach the slope parameter $\alpha$ is not a fit parameter but it is fixed to a value
that improves the angular resolution.
\item[{\bf c)}] {\bf New planar fit}\\
With the time values $t'_i$ we reconstruct a new shower direction by means of a further planar fit.
\end{itemize}
In Fig. \ref{residuo1} the reduction of the non-gaussian tailes as a function of the 
successive plane fit iterations is shown. 
After few iterations the time residuals distribution assume a symmetrical shape 
(dotted histograms).

In order to optimize this procedure we have studied the resulting angular resolution as a function
of the number of planar fit iterations and of the values for $K$.
To estimate the performance of the algorithms we have used the $\psi_{70}$ parameter, defined
as the value in the distribution of the angle between the true and the reconstructed
directions within which 71.5 $\%$ of the events are contained. 
The angular resolution $\sigma_{\theta}$ is given by $\sigma_{\theta} = 1.58 \cdot \psi_{70}$.
\begin{figure}[t]
\begin{minipage}[t]{.48\linewidth}
   \vspace{5.8cm}
\includegraphics{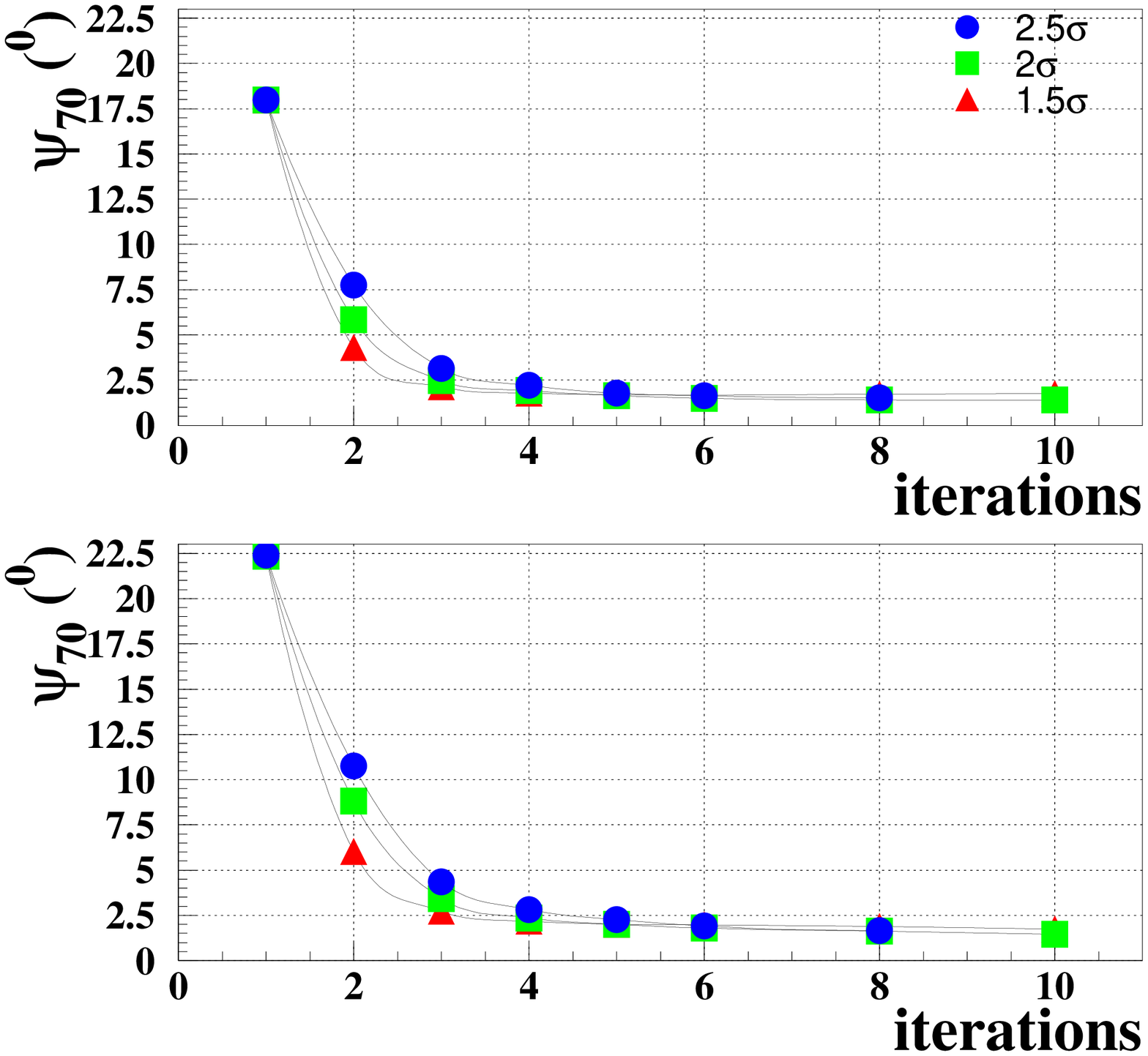}
 \caption{\it Opening angle $\psi_{70}$ vs. number of iterations for different values  
of the $K$ parameter. The upper plot refers to 1 TeV $\gamma$-induced showers, 
the lower one to 500 GeV ($N_{hit} = 20 - 50$).
    \label{psi_iter1} }
 \end{minipage}\hfill     
\begin{minipage}[t]{.48\linewidth}
   \vspace{5.8cm}
    \includegraphics{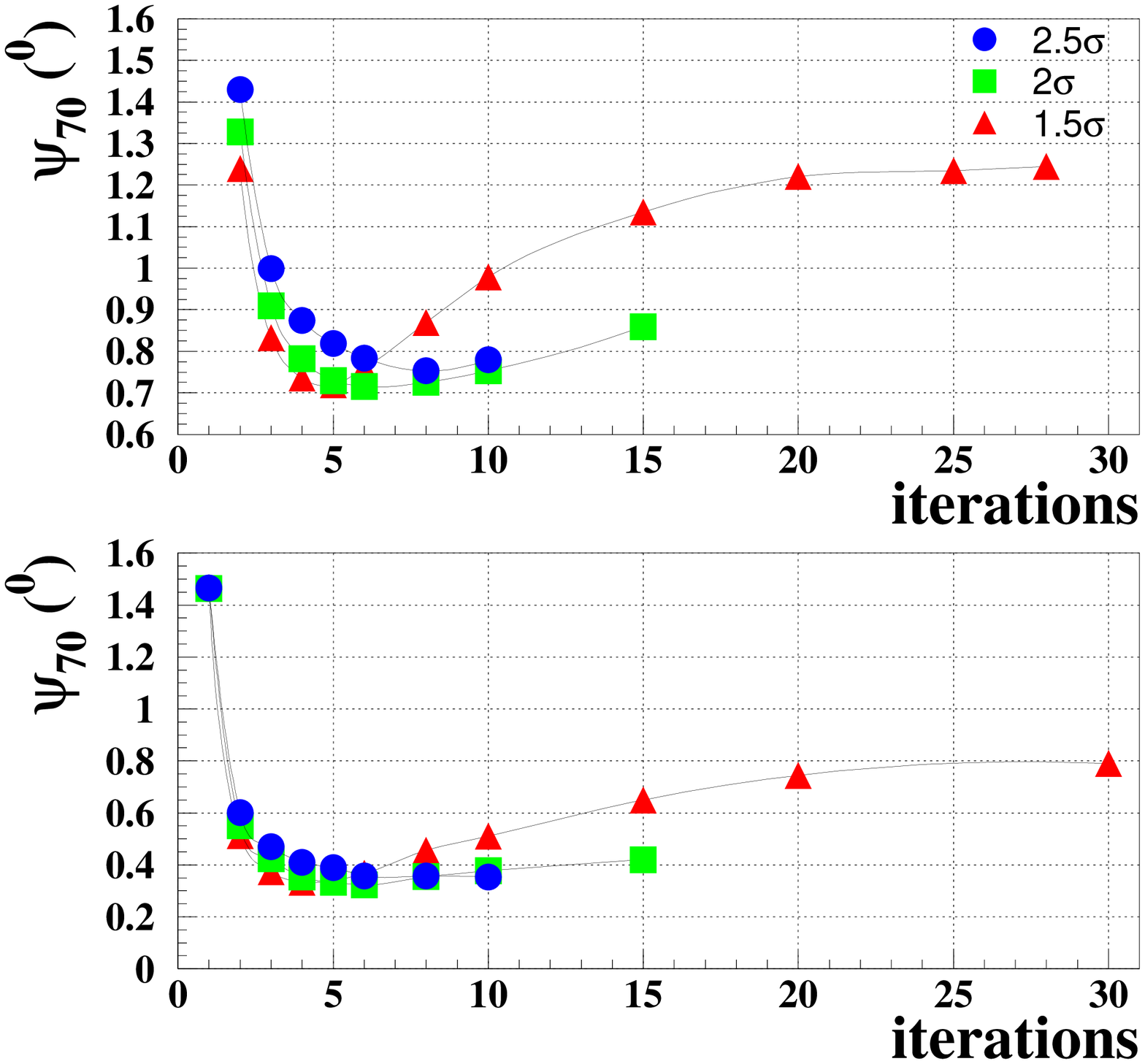}
    \caption{\it Opening angle $\psi_{70}$ vs. number of iterations for different values  
of the $K$ parameter. The upper plot refers to 1 TeV $\gamma$-induced showers
($N_{hit} = 100 - 200$), the lower one to 5 TeV ($N_{hit} = 500 - 600$).
 \label{psi_iter2} }
 \end{minipage}\hfill        
\end{figure}
In Fig. \ref{psi_iter1} the behaviour of the opening angle $\psi_{70}$ as a 
function of the number of iterations is shown, for different choices of $K$. The upper plot
refers to 1 TeV $\gamma$-induced showers, the lower to 500 GeV $\gamma$-induced showers.
In these calculations, all showers have been sampled with the core at the center of the carpet
and we have selected events with a pad multiplicity $N_{hit}$ in the range 20 - 50.
The same plots for higher multiplicities are shown is Fig. \ref{psi_iter2}.
No converter has been added on the RPCs.

From these figures it results:
\begin{itemize}
\item The angular resolution improves substantially with successive iterations. 
Indeed, the successive minimizations, as expected, reject the farthest temporal hits from 
the ``theoretical plane'' (see Fig. \ref{residuo1}).
This behaviour is qualitatively independent of the primary energy, the pad multiplicity and 
the values of $K$.
\item After a few iterations the improvement stops and, for high multiplicities, 
the angular resolution worsens (see Fig. \ref{psi_iter2}).
This feature is due to the large fraction of discarded temporal hits lying
far from the shower core. As a consequence, we must fix a suitable maximum number of iterations.
\end{itemize}
A systematic study\cite{elly} performed for different energies and zenithal angles, 
sampling the shower 
cores over a wide area, led us to the following best tuning of the planar fit procedure:
$K=1.5 - 2$ with the maximum number of iterations in the range 4 - 6.

\begin{figure}[t]
\begin{minipage}[t]{.48\linewidth}
   \vspace{5.8cm}
\includegraphics{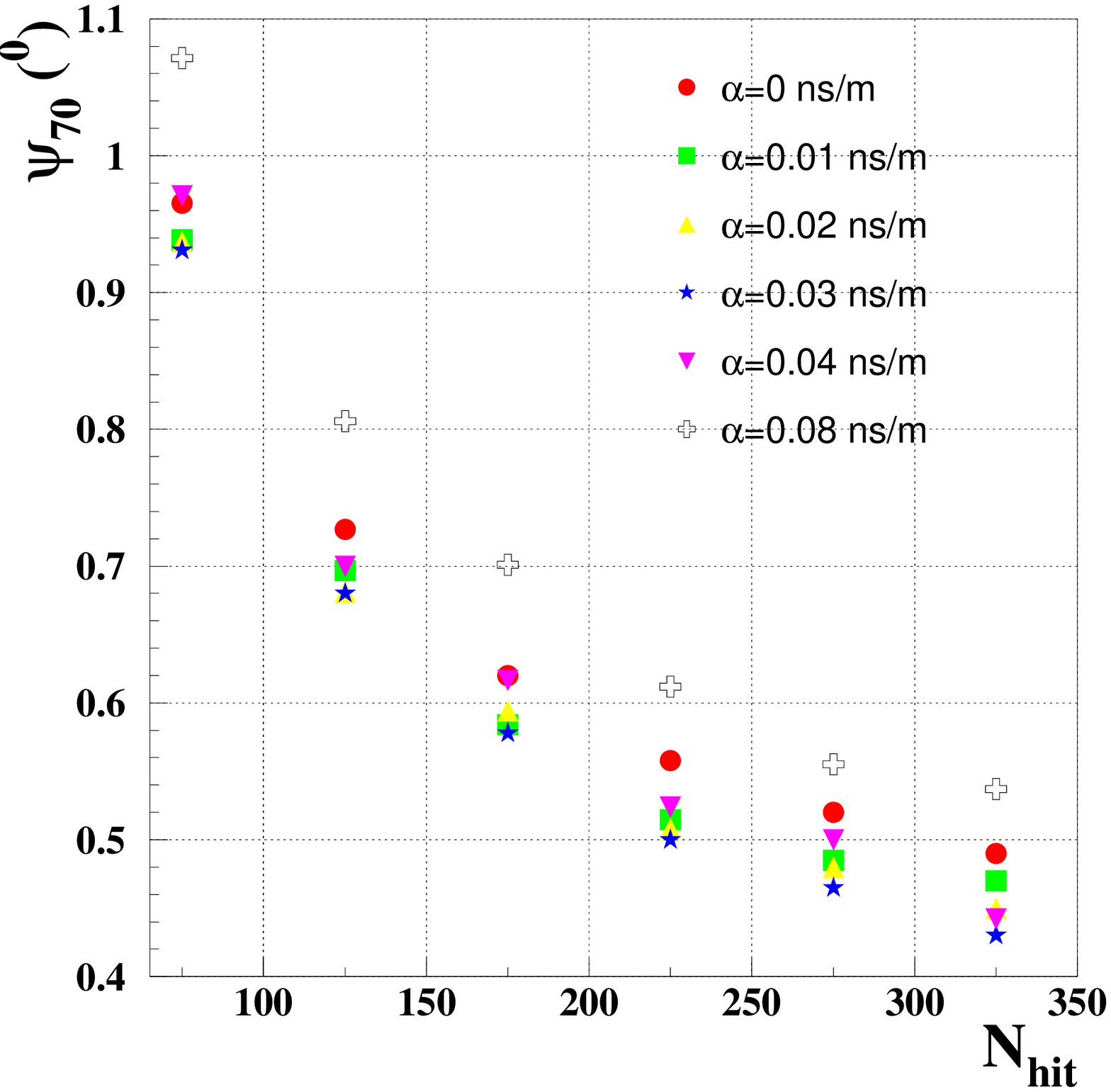}
 \caption{\it The parameter $\psi_{70}$ as a function of pad multiplicity for different values 
of the conical parameter $\alpha$ in vertical $1~TeV$ $\gamma$-induced showers.
    \label{fig415} }
 \end{minipage}\hfill     
\begin{minipage}[t]{.48\linewidth}
   \vspace{5.8cm}
    \includegraphics{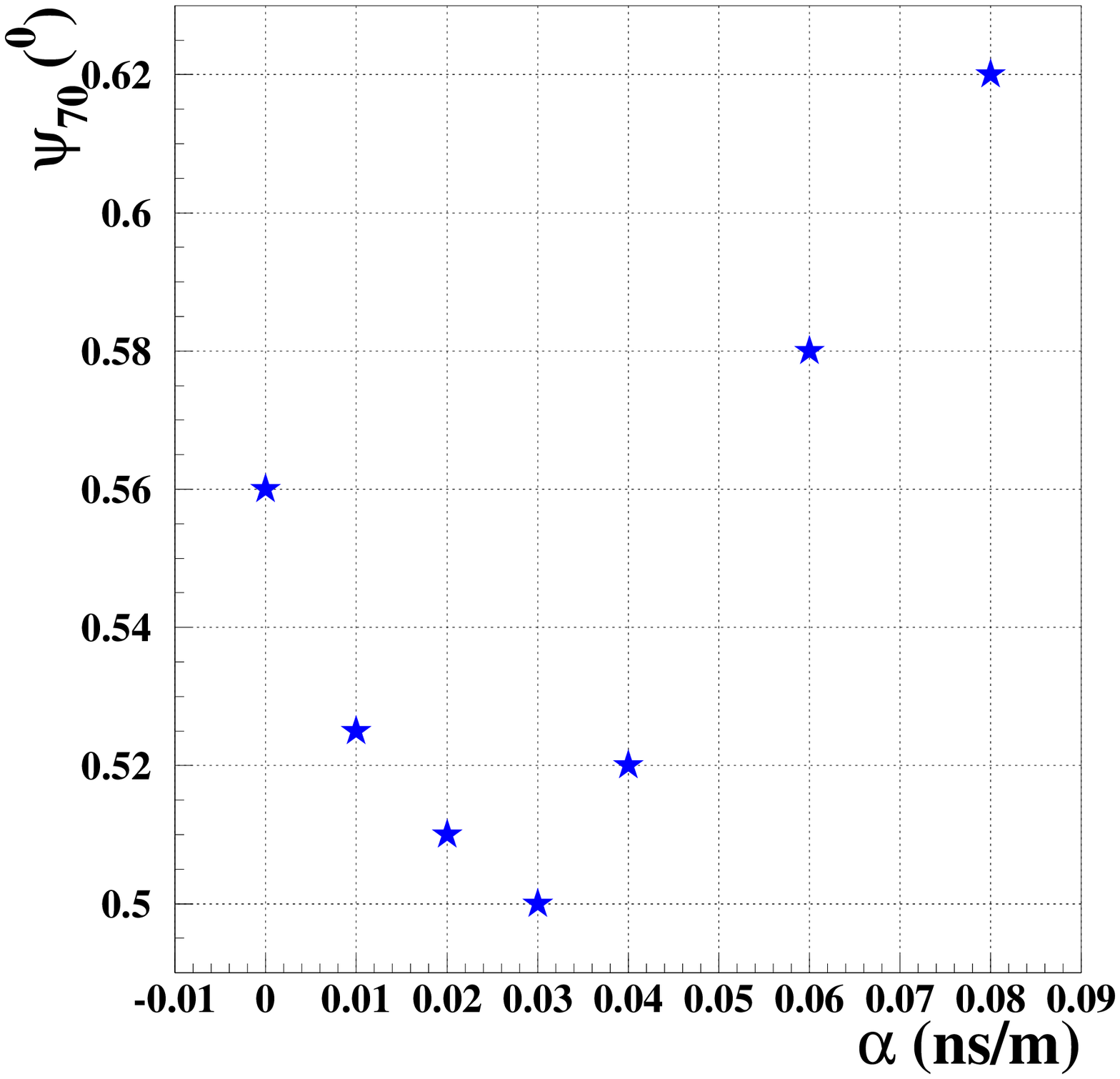}
    \caption{\it Opening angle $\psi_{70}$ as a function of the conical parameter $\alpha$ for
vertical $1~TeV$ $\gamma$-induced showers with a selected pad multiplicity $N_{hit}= 200-250$. 
 \label{alpha} }
 \end{minipage}\hfill        
\end{figure}

A further improvement of the angular resolution can be obtained by fixing the cone slope of the
successive conical correction to the value $\alpha = 0.03$ $ns/m$.
As an example, in Fig. \ref{fig415} we show the parameter $\psi_{70}$ as a function of pad 
multiplicity for different values of the cone slope $\alpha$. The calculations refer to 
vertical 1 TeV $\gamma$-induced showers with the core at the center of the carpet.
As can be seen, the value  $\alpha= 0.03$ $ns/m$ improves the angular resolution for all the
investigated multiplicities (see Fig. \ref{alpha}). 

We also investigated the performance of a conical fit in which the cone slope is a free 
parameter of the minimization\cite{elly}. 
The angular resolution calculated with fixed $\alpha$ gives better results because
the minimization tends to adapte the cone slope to the temporal hits, thus introducing
a larger error in the direction recontruction for events with the core outside the carpet
or in its boundaries\cite{icrc}.

\subsection{The effect of a lead converter on the angular resolution}

The consequences of placing a thin sheet of converter above the detector are, qualitatively:
(1) absorbtion of low energy electrons (and photons) which no longer contribute to the time signal;
(2) multiplication process of high-energy electrons and photons which produces a signal
enhancement. The lead converter thus reduces the temporal fluctuations: the contributions gained
are concentrated near the ideal time profile since the high energy particles travel near the
shower front while those lost tend to lag far behind.

In order to study the effect of the converter on the angular resolution we have simulated, 
via a GEANT3-based code, a 0.5 cm lead sheet located 5 cm above the RPCs.
The results discussed in the previous section are qualitatively reproduced also when the 
converter is added for what concerns the dependence of the angular resolution on the number 
of iterations of the planar fit.
As expected, the improvement of the showers temporal profile implies a fewer number of
iterations in the iterative procedure: the best tuning of the planar fit is achieved after only 
3 iterations, with K = 1.5 .
\begin{figure}[t]
\begin{minipage}[t]{.48\linewidth}
   \vspace{5.8cm}
\includegraphics{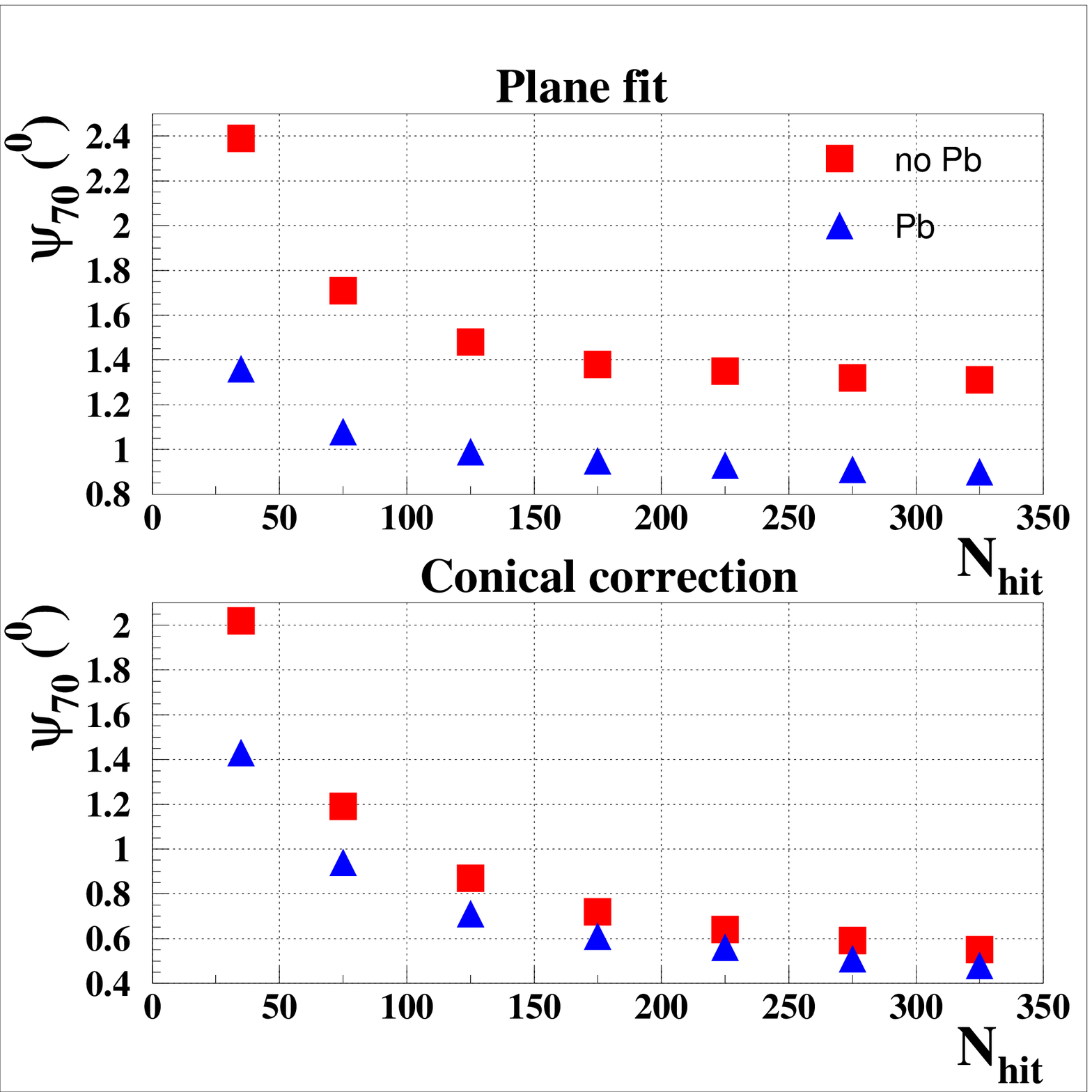}
 \caption{\it The parameter $\psi_{70}$ as a function of pad multiplicity after the planar fit 
(upper plot) and after the conical correction (lower plot) for $1~TeV$ $\gamma$-induced 
showers ($\theta=20^{\circ}$). 
    \label{plan_con} }
 \end{minipage}\hfill     
\begin{minipage}[t]{.48\linewidth}
   \vspace{5.8cm}
    \includegraphics{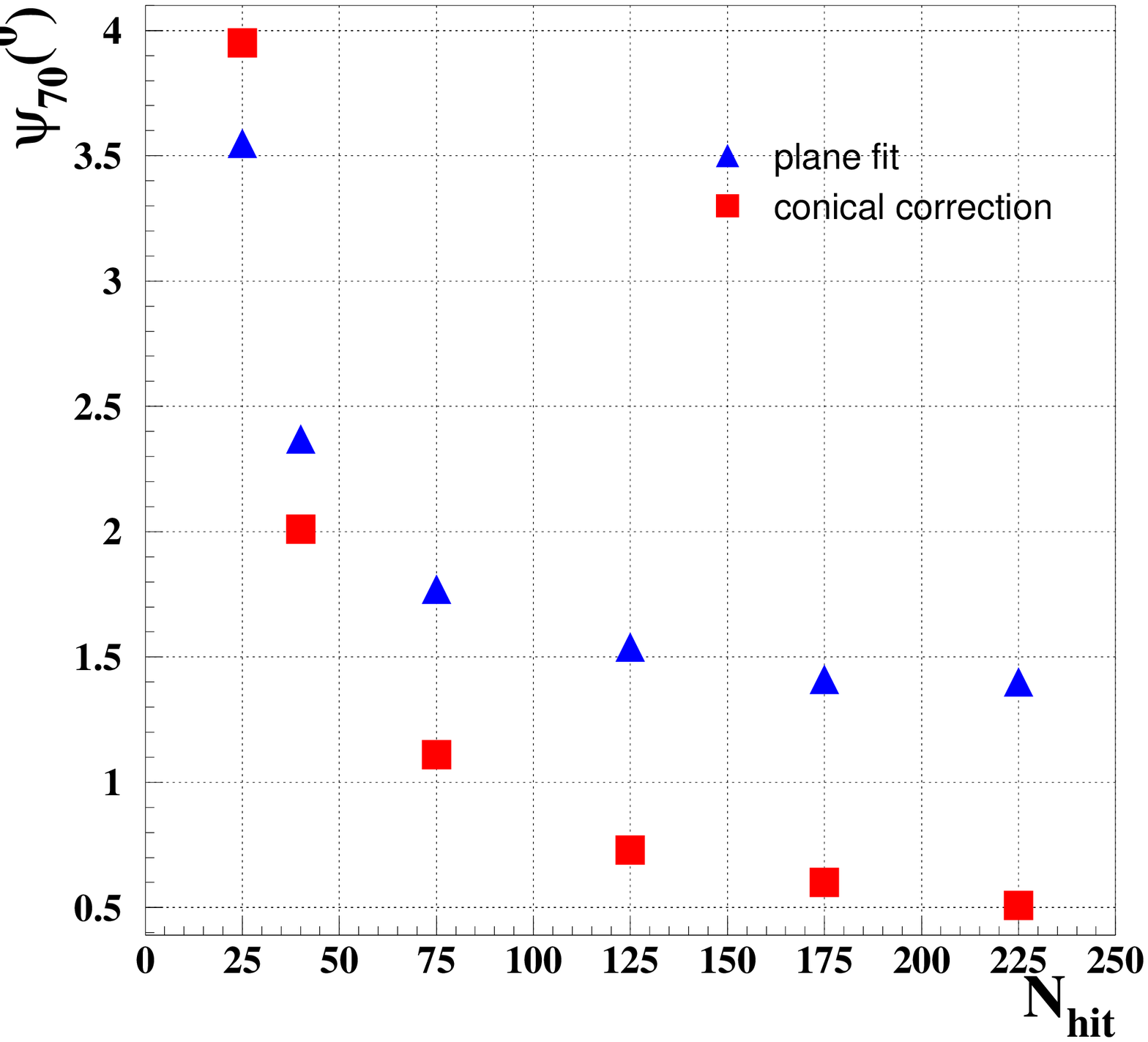}
    \caption{\it The parameter $\psi_{70}$ as a function of pad multiplicity for $\gamma$-induced
showers. We compare the opening angle calculated with different algorithms.
 \label{psi70_piano} }
 \end{minipage}\hfill        
\end{figure}
In Fig. \ref{plan_con} we show the effect of a 0.5 cm lead sheet on the angular resolution by 
comparing the $\psi_{70}$ parameter as a function of the fired pads with and without the lead.
The calculations refer to 1 TeV $\gamma$-induced showers with a fixed zenithal angle 
$\theta=20^{\circ}$ and with the core sampled on a 50 $\times$ 50 m$^2$ area. 
In the upper plot we show the values calculated after the best tuned iterative plane fit: 
5 iterations without lead and 3 iterations with the converter.
In the lower part of the figure $\psi_{70}$ is calculated after the conical correction. 
The improvement is clearly evident.


\section{Angular resolution of the ARGO-YBJ detector}

In oder to study the angular resolution of the ARGO-YBJ detector we have simulated, via the 
Corsika code\cite{corsika}, $\gamma$-induced showers with a Crab-like spectrum ($\sim~E^{-2.5}$) and 
proton events with $\sim~E^{-2.75}$, both ranging from 100 GeV to 50 TeV. The $\gamma$-rays have been simulated for different zenith angles ($<40^{\circ}$), following the daily path of the source in the sky. The detector response has been simulated via a GEANT3-based code.
A 0.5 cm lead sheet have been considered 5 cm above the RPCs.
The core positions have been randomly sampled in an area, energy-dependent, large up to 
800 $\times$ 800 m$^2$ and centered on the detector. 

In Fig. \ref{psi70_piano} we compare the dependence of the opening angle $\psi_{70}$ on 
pad multiplicity for the different algorithms discussed above. The $\gamma$-induced showers 
are reconstructed inside a fiducial area similar to the carpet dimensions
($A_{fid}= 80 \times 80$ $m^2$). 
The shower core positions are reconstructed by means of the Maximum Likelihood Method applied
on the RPCs (the so-called 'LLF2 method'\cite{inout}).
The conical correction needs a good shower core position resolution, so for very low multiplicities the planar fit is more performant because of the high  contamination of external events.
However, when suitable ``cuts''\cite{inout} 
are applied and a high fraction of external showers is rejected, the conical 
correction improves the planar fit.
\begin{figure}
\begin{minipage}[t]{.48\linewidth}
   \vspace{5.8cm}
    \includegraphics{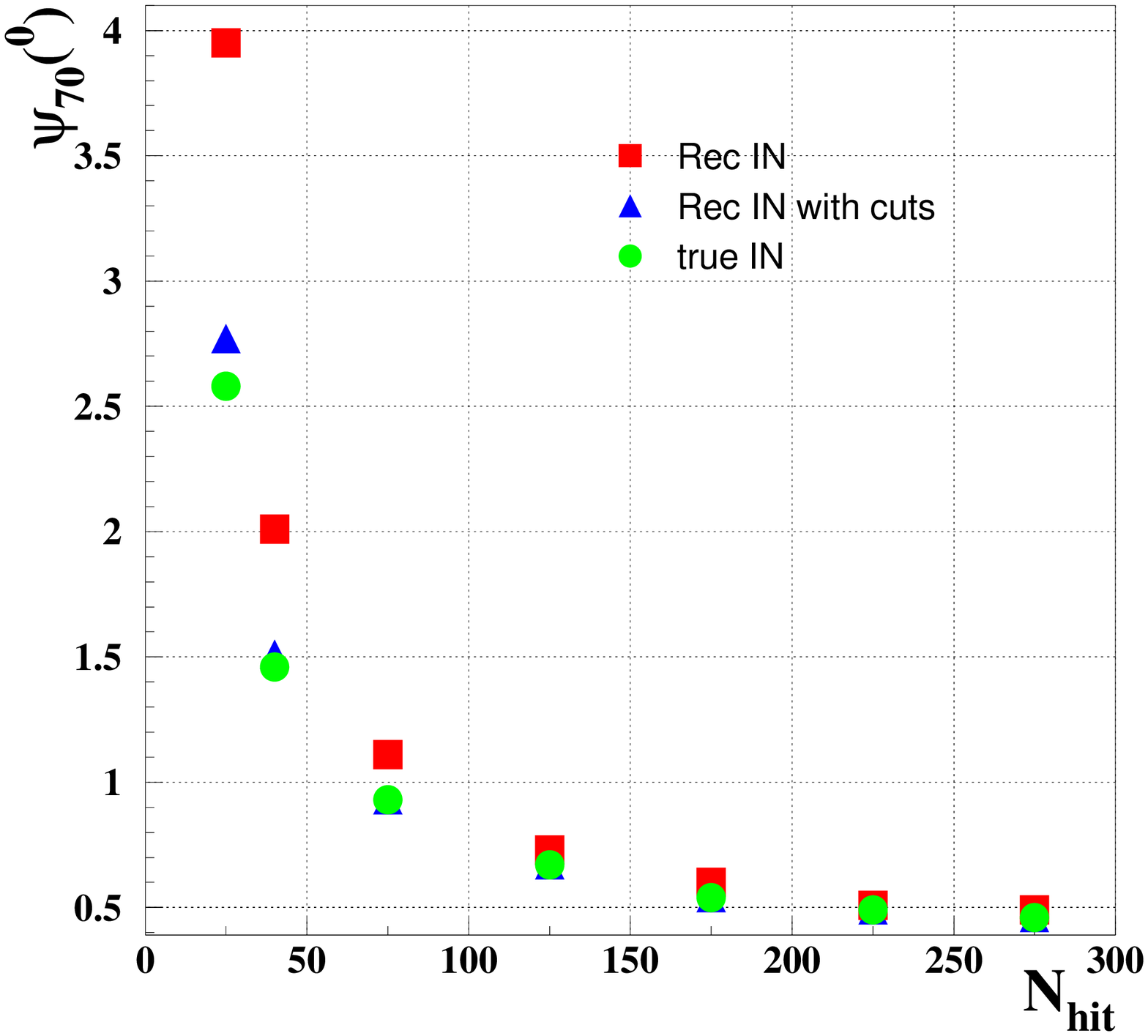}
    \caption{\it The parameter $\psi_{70}$ as a function of pad multiplicity 
for $\gamma$-induced showers after the conical correction. 
 \label{psi70_g} }
 \end{minipage}\hfill        
\begin{minipage}[t]{.48\linewidth}
   \vspace{5.8cm}
    \includegraphics{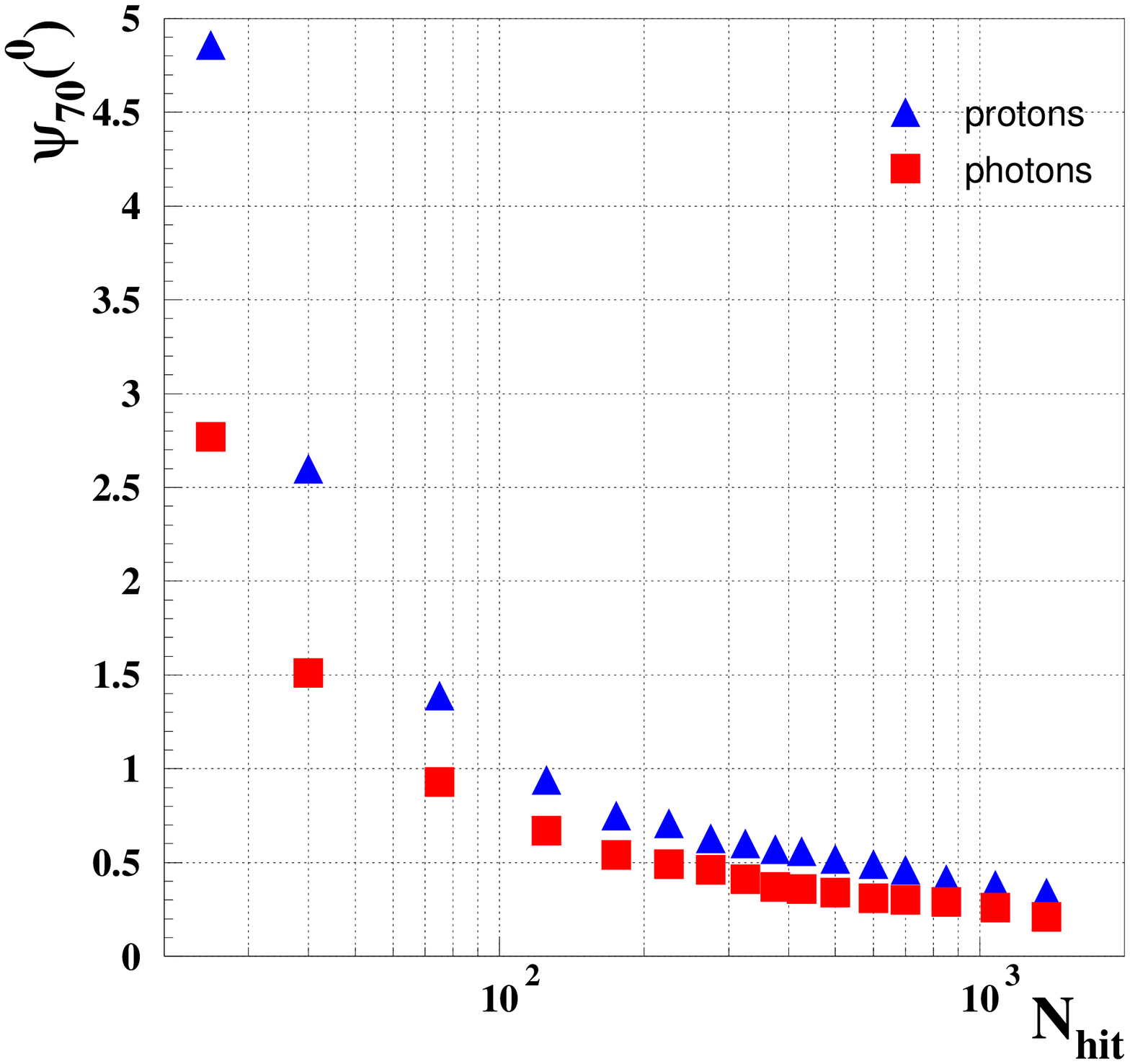}
    \caption{\it The parameter $\psi_{70}$ as a function of pad multiplicity for $\gamma$ 
and proton showers. 
 \label{psi70_gp} }
 \end{minipage}\hfill        
\end{figure}
In Fig. \ref{psi70_g} we show the parameter $\psi_{70}$ as a 
function of pad multiplicity for $\gamma$ showers reconstructed inside $A_{fid}$ with and 
without the application of the cuts. 
For comparison, the $\psi_{70}$ values for the events with the sampled core truly inside $A_{fid}$ 
are also plotted (circles). 
As can be seen, for low multiplicities ($N_{hit} < 100$), the rejection of external events is
crucial to improve the angular resolution.

In Fig. \ref{psi70_gp} a comparison between the angular resolutions obtained after the cuts for gamma
and proton showers is shown. Due to the larger lateral spread of particles in proton showers, 
the angular resolution for protons is worse than that for photons.

\section{Conclusions}

In this paper we described an algorithm optimized to reconstruct the direction of the showers
detected by the ARGO-YBJ experiment. A conical correction with a cone slope
parameter fixed to the value $\alpha=0.03$ $ns/m$ is proposed and discussed.
The resulting opening angle $\psi_{70}$ is better than $0.3^{\circ}$ for $\gamma$-induced events 
which fire more than 600 pads on the carpet.

For $N_{hit} > 100$ (when the dependence on the fiducial area is reduced) the angular resolution 
of the ARGO-YBJ experiment, for $\gamma$-induced events with a Crab-like energy spectra, 
can be described by the equation $\psi_{70} = 0.046 + 6.666/ \sqrt{N_{hit}}$.



\end{document}